\documentclass[aps,graphicx,reprint,amsmath,amssymb,prb]{revtex4-1}
\usepackage{amsfonts}
\usepackage{}
\usepackage{graphicx}
\usepackage{color}
\begin{document}
\title{Many-body effects on electron spin dynamics in semiconductors from a geometrical viewpoint}
\author{Chunbo Zhao}

\email[]{cbzhao@semi.ac.cn}
\affiliation{State Key Laboratory for Superlattices and Microstructures, Institute of Semiconductors\\Chinese Academy of Sciences, P.O.Box 912,Beijing 100083, People's Republic of China}
\date{\today}
\begin{abstract}
  Many body effects on spin dynamics in semiconductors have attracted a lot of attentions in recent years. In this paper, we show why and how the many body effects have to be considered by a simple Bloch sphere geometry approach. The micro dynamics here are viewed as a time dependent sequence of unitary group action on the spin density matrix. Based on this physical picture, we give the explicit unitary group for conventional spin dynamics mechanism such as DP, EY, and BAP using pure density matrix.  And we also show the various scattering processes how influence the spin system via mixed density matrix and Feynman diagrams.
 \end{abstract}
\maketitle
\section{Introduction}
Semiconductor spintronics, which aims at utilizing or incorporating the spin degree of freedom in electronics, has attracted great interests in last decades years\cite{meier1984optical,vzutic2004spintronics,awschalom2002semiconductor}. Many novel spin-related phenomena and properties, such as the spin Hall effect\cite{kato2004observation,wunderlich2004experimental}, spin Coulomb drag effect\cite{d2000theory,weber2005observation}, spin photogavanic effect\cite{ganichev2000circular,ganichev2002spin} and persistent spin helix effect\cite{koralek2009emergence,bernevig2006exact}, have been discovered.  These novel spin-related physics can be partly understood well by single-particle demonstration. However, the real physical system is interacting, there may be something new physics emerging only when considering many-body effects. In fact, recently Wu, el\cite{wu2010spin} developed a fully microscopic many-body theory on spin dynamics in semiconductors called kinetic spin Bloch equation (KSBE) using non-equilibrium Green function\cite{haug2008quantum}, considering different kinds of scattering. From this theory, they predicted many novel effects, such as nonmontonic spin relaxation time dependence of temperature or electron density in GaAs quantum well\cite{zhou2007spin}, hole screening effect in hole doped bulk $p$-GaAs\cite{jiang2009electron} and so on. And later these predictions have been observed experimentally one after another by time-resolved Kerr rotation technique\cite{han2011temperature,zhao2013electron}. Even though KSBE is useful, one cannot easily get the underlying physics immediately from the numeric results of KSBE since there is no analytical result generally for its complicated expression.  Encouraged by the powerful of KSBE, it will be helpful and profound to reinterpret the many body effects on spin dynamics from another viewpoint if this interpretation can give us a clear and insightful to the questions addressed.
\par

In this paper, we try to give an intuitive picture about the many-body effects on spin dynamics using the Bloch sphere geometry language. This demonstration will give us a general framework to interpret the explicit dynamics emerging from spin-orbit coupling or/and many body scattering, which will provide us a geometrical way to understand the numeric results of KSBE.  The main idea is that we view the spin dynamics as a time-dependent unitary group element of $SU(2)$ action on the spin density matrix. Based on this picture, the electron spin dynamics can be understood as a vector rotation in Bloch sphere if one applies the mathematical map called Hopf map. Therefore, if we denote the spin 'direction' as a point on the Bloch sphere,  then the resultant spin 'direction' will be the point after a $SO(3)$ group action. Here we want to mention that: even though the Hopf map is not one-to-one, the Bloch sphere interpretation of spin dynamics captures the essential physics process that we investigated. The rest paper is organized as follows. Firstly, we present the pure and mixed density matrix with a unitary group. Secondly, we will try to transform the well studied spin dynamics mechanisms such as D'yakonov-Perel (DP), Elliott-Yafet (EY), and Bir-Aronov-Pikus (BAP) to a unitary element in $SU(2)$. Finally, we will use the mixed spin density matrix and Fennyman diagram to account for the many body effects with different scattering considered.
\section{spin dynamics in semiconductors}
\subsection{pure and mixed spin density matrix}
It is known that the single electron spin space can be expressed with a two components complex-valued function
$|\phi\rangle=\left(\begin{array}{cc} z_1 , & z_2\end{array}\right){}^T$. If we suppose  the normalizing condition as
\begin{equation}\label{spin}
    \langle\phi|\phi\rangle=z_1^*z_1+z_2^*z_2=1,
\end{equation}
 then the electron spin space is nothing but a sphere $S^3$  in the 4-d Euclid space with symmetry group $SU(2)$.  Physically, the spin state or wave function $|\phi\rangle $ will be changed under the external magnetic field or electric field by Rashaba spin-orbit coupling. Mathematically, the dynamics of wave function $|\phi\rangle$ can be interpreted as a $SU(2)$ group element $u$ acting on the former wave funtion $|\phi^{'}\rangle$, which can be regarded as just a rotation of wave function under the symmetry group $u$, $|\phi\rangle =u|\phi^{'}\rangle$. For the convenience of the description, one usually apply the spin density operator $\rho$ (which is isomorphism to the space of wavefunction) to study spin dynamics. For a pure state, the density matrix is expressed as
\begin{equation}\label{2}
    \rho=|\phi\rangle\langle\phi|=
\left(
\begin{array}{cc}
 z_1z_1^* & z_1z_2^* \\
 z_2z_1^* & z_2z_2^*
\end{array}
\right)
\end{equation}
 According to this representation, one can obtain the one-to-one map through $\langle\phi|\phi\rangle\leftrightarrow|\phi\rangle\langle\phi|$ between wave function and density matrix space. Hence, we can investigate the spin dynamics by density matrix or its matric representations of SU(2). Due to any element of SU(2) can be decomposed using Pauli matrices set and identity operator,  we have a compact form of the density matrix (\ref{2}) written in terms of $2\times2 $ Pauli matrices set $\{\mathbb{I}, \sigma_x, \sigma_y, \sigma_z\}$,
 \begin{equation}\label{3}
    \rho=x_0\mathbb{I}+x_1\sigma_x+x_2\sigma_y+x_3\sigma_z,
 \end{equation}
 where $x_0, x_1, x_2, x_3$ are real numbers, $\mathbb{I}$ is the identity matrix.
Applying the Eq.(\ref{spin}), a constraint of density matrix can be obtained:
 \begin{equation}\label{5}
    tr\rho=z_1^*z_1+z_2^*z_2=2x_0=1
 \end{equation}
 so $x_0=1/2$ is required. We rewrite the density matrix as
 \begin{equation}\label{6}
    \rho=\frac{1}{2}(I+\vec{x}\cdot\vec{\sigma})
 \end{equation}
 where $\vec{x}=(x_1,x_2,x_3)$ and $\vec{\sigma}=(\sigma_x, \sigma_y, \sigma_z)$ is the vector with the three Pauli matrices as components. As for pure state, $tr(\rho^2)=1$\cite{blum2012density}, which leads to the constraint of $|\vec{x}|^2 = x_1^2+x_2^2+x_3^2 = 1$. Hence, a point in $S^2$ can be a presentation of $\vec{x}$. This map of $SU(2)$ to $SO(3)$ is nothing but the Hopf map $S^3\rightarrow S^2$ in mathematics.
\par
 So far, we have introduced the pure state density matrix, next we will discuss the more realistic case, called mixed state matrix, in which, the many body effects will be emerged naturally.  Since an electron quantum state in semiconductor can be labeled by  momentum $\vec{k}$ and energy band index $n$, so we can denote different state electrons by quantum number $\vec{k}$ when only conduction band is considered. The whole density matrix can be written as follows\cite{blum2012density},
\begin{equation}\label{mixed spin density matrix}
    \rho=\sum_{\vec{k}}|\phi_{\vec{k}}\rangle p_{\vec{k}} \langle\phi_{\vec{k}}|
\end{equation}
where $\sum_{\vec{k}} p_{\vec{k}}=1$, $|\phi_{\vec{k}}\rangle$ is the spin space of state with momentum $\textbf{k}$ , $p_{\vec{k}}$ is actually its corresponding probability (statistical weights). Hence,  for a physical operator $A$, the measured quantity will need to be averaged twice, one is the quantum mechanics average, the second is the statistical average, it can be described as:
\begin{equation}\label{physical quantity}
    \langle\langle A\rangle\rangle=\sum_{\vec{k}} p_{\vec{k}} \langle \phi_{\vec{k}}|A|\phi_{\vec{k}}\rangle=\textrm{tr}(A\rho),
\end{equation}
where $\textrm{tr}$ is the trace operator. For convenience later, we define single state density matrix as $\rho_{\vec{k}}=|\phi_{\vec{k}}\rangle p_{\vec{k}} \langle\phi_{\vec{k}}|$ for state labeled with $\vec{k}$.
 Eq.(\ref{physical quantity}) clearly indicates that the probability $p_{\vec{k}}$ containing in $\rho_{\vec{k}}$ will influnce the physical observations. Therefore, scattering processes such as electron-electron, electron-phonon, electron-impurity scatterings, will influence the physical quantity since the state at $\vec{k}$ will be scattered to the one at $\vec{k^{'}}$, but they may not have the equal statistical probability ($p_{\vec{k}}\neq p_{\vec{k'}}$). There are two important properties for mixed state density matrix, one is that the trace of $\rho$ is identity $\textrm{tr}(\rho)=\sum_k p_k=1$, which is the same with pure state. The other special property is that $\textrm{tr}(\rho^2) < \sum_{\vec{k}} p_{\vec{k}}=1$, which is the only feature of the mixed state.
 \par
 If we know the original physical state, assuming the spin direction pointing to $\textbf{z}$ axis, then one can follow the tracks of spin dynamics via Eq.(\ref{physical quantity}), $S_z(t) = \textrm{tr}(\sigma_z \rho(t))$. So the main task is to study the evolution of mixed density matrix $\rho$. In the following part, we present how the dynamic process can be viewed as a simple vector rotation in Bloch sphere for single and many body effects on electron spin dynamics. And we also give the physical picture to understand the KSBE applying our method.
\subsection{spin relaxation mechanisms}

 \par
\textcolor[rgb]{0.00,0.00,0.00}{\textbf{D'yakonov-Perel's mechanism (DP)}}
 \par
\begin{figure}[h]
  \scalebox{0.4}[0.4]{\includegraphics{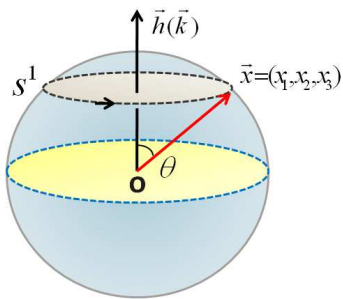}}\\
  \caption{ The spin direction denoted with vector $\vec{x}$ rotates around the effective magnetic field $\vec{h}$, which can be used to demonstrate the Dresshauls and Rashaba effects on spin relaxation during the scattering.}\label{larmor}
\end{figure}

 According to the Liouville equation of density operator\cite{blum2012density}, we have the following motion equation for the density matrix $\rho$ :

 \begin{equation}\label{4}
  \dot{\textbf{$\rho$}}=-i[\textbf{$H$},\textbf{$\rho$}],
 \end{equation}
 where $[~~ ]$ is the commutator operator, and here we assume $\hbar=1$ for simplicity.
 For example, when the hamitonian $H$ can be written as Dreeslhaus and Rashaba form,
$H=\gamma(\Omega_D(\vec{k})+\Omega_R(\vec{k}))\cdot\vec{\sigma}=\frac{1}{2}\vec{h}(\vec{k})\cdot\vec{\sigma}$, the motion of $\rho$ describes the respective scattering process:
\begin{equation}\label{7}
    \dot{\textbf{$\rho$}}=-i[\gamma(\Omega_D(\vec{k})+\Omega_R(\vec{k}))\cdot\vec{\sigma},\textbf{$\rho$}]
    =-i[\frac{1}{2}h(\vec{k})\cdot\vec{\sigma},\textbf{$\rho$}]
\end{equation}
Substitute Eq.(\ref{6}) into the Eq.(\ref{7}), one can easily obtain the following:
\begin{equation}\label{8}
    \dot{\vec{x}}=\vec{h}(\vec{k})\times\vec{x},
\end{equation}
where we have used the identity $(\vec{\sigma}\cdot\vec{a})(\vec{\sigma}\cdot\vec{b})=\mathbb{I}
(\vec{a}\cdot\vec{b})+i\vec{\sigma}\cdot(\vec{a}\times\vec{b})$
for arbitrary vector $\vec{a}$ and $\vec{b}$.  Actually, we can use a unitary group element to reexpress this motion as $\rho^{'}=u(\vec{h}(t))\rho$, where
\begin{equation}\label{uDP}
    u(\vec{h}(t))=
\exp(-i(\vec{h}(t)\cdot\vec{\sigma})/2).
\end{equation}
Thus the temporal evolution of density matrix can be understood as a unitary element $u(\vec{h}(t))$ acting on the previous one. If we apply the Hopf map, this process can be described by Eq.(\ref{8}), which is nothing but the familiar Larmor precession equation. It can be intuitively interpreted by Fig.\ref{larmor}.

\par
The electron spin denoted as $\vec{x}$ will precess around the randomly distributed effective  magnetic filed $\vec{h}(\vec{k})$ during the interval of scattering. Since the process can be viewed as a free electron motion during this time, it's during this time that the Hamiltonian has the form discussed above. In the presence of momentum scattering, electron changes its momentum $\vec{k}$ randomly, hence spins precess randomly between the adjacent scattering events. This random-walk-like evolution of spin phase leads to spin relaxation, that is, the so called the DP mechanism\cite{d1971spin,d1972spin}.

\par
\textcolor[rgb]{0.00,0.00,0.00}{\textbf{Elliott-Yafet mechanism (EY)}}

\par
\begin{figure}[h]
  \scalebox{0.4}[0.4]{\includegraphics{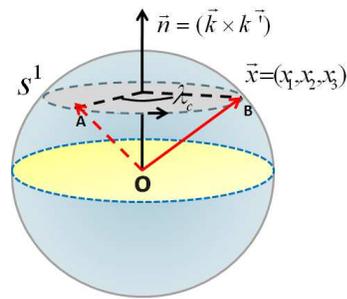}}\\
  \caption{ The $\vec{n}$ is the rotation axis of the spin direction $\vec{x}$, $\lambda_c$ is the angle of rotation about the axis $\vec{n}$, during every scattering process, the spin-independent scattering will lead to additional spin-flip because this process proposed by Elliott and Yafet, this physical picture can be described in this Bloch sphere geometry like this figure. After the scattering, the spin will be rotated an angle $\lambda_c$ which is related to specific materials. Above picture clearly indicates the flip effect of EY mechanism. }\label{EY}
\end{figure}

 Another important spin relaxation mechanism is called EY mechanism\cite{elliott1954theory,yafet1963g}. It was pointed out by Elliott that the spin-up and spin-down electronic eigenstates states mix due to the spin-orbit interaction, so the spin direction will be flipped after the scattering. Hence different kinds of spin-independent scattering can cause spin flip and thereby affect the spin relaxation process. The spin-flip matrix was usually written approximately as follows\cite{pikus1984spin}:
\begin{equation}\label{13}
  \Lambda_{\vec{k},\vec{k'}}=\mathbb{I}-i\lambda_c[(\vec{k}\times\vec{k'})\cdot\vec{\sigma}]
\end{equation}
where $\lambda_c$ is parameters of the studied materials\cite{wu2010spin}. However, one can describe this phenomenon geometrically as follows:
\begin{eqnarray}\label{14}
  \Lambda_{\vec{k},\vec{k'}}\simeq u(\vec{n},\vec{\omega})
   \nonumber&=& \mathbb{I}\cos(\omega/2)-i(\vec{n}\cdot\vec{\sigma})\sin(\omega/2) \\
            &=&   \exp(-i(\vec{n}\cdot\vec{\sigma})\omega/2)
\end{eqnarray}
where $\vec{n}=\vec{k}\times\vec{k'}$, and we assume $\omega=2\lambda_c$.  Eq.(\ref{14}) indicates  that $u(\vec{n},\vec{\omega})\in SU(2)$, and therefore the density matrix can be changed when EY process occurs, such as $\rho^{'}=u(\vec{n},\vec{\omega})\rho$.  Hence, when two electron scattering happens, we have to insert two unitary matrices $\Lambda_{\vec{k},\vec{k'}} (\Lambda_{\vec{k'},\vec{k}})$ before the density matrix $\rho_{\vec{k'}} (\rho_{\vec{k}})$ if considering the EY  mechanism\cite{jiang2009electron}. Based on the interpretation talked above, we can give the spin dynamics process like in Fig.\ref{EY}, where electron spin direction $\vec{x}$ of state $\vec{k}$ will rotate an angle $\lambda_c$ from position A to B as denoted  in the sphere.
\par
\textcolor[rgb]{0.00,0.00,0.00}{\textbf{Bir-Aronov-Pikus mechanism (BAP)}}\par
It was proposed by Bir, Aronov and Pikus that the electron-hole exchange scattering can lead to efficient electron spin relaxation in $p$-type semiconductors\cite{aronov1983spin}. Physically, BAP spin mechanism results from the repulsion of Coulomb force and  antisymmetric property of wavefunction.  Considering both short and long range part of hole-electron exchange interaction, the  general Hamiltonian can be written as a compact form \cite{maialle1996spin,jiang2009electron}
\begin{equation}\label{bap}
    H_{ex}=\delta_{\vec{K},\vec{K'}}\hat{\mathcal {J}}\cdot\hat{S},
\end{equation}
where $\vec{K}=\vec{k}_e+\vec{k}_h$ is the sum of electron and hole wavevectors which participate in the interaction. The operator of $\hat{\mathcal{J}}$ is a $4\times4$ matrix in hole spin space, $\hat{S}$ is the electron spin operator (also 1/2 Pauli matrix), they can rotate the hole and electron spin directions, respectively.  Since the exchange coupling term of Eq.(\ref{bap}), the variation of hole spin direction will affect the electron spin direction correspondingly.  Mathematically, the hole spin space is equivalent to a unit sphere in 8-d Euclid space, which can be demonstrated using four-components complex-valued wave function $|\phi\rangle=\left(\begin{array}{cc} z_1,z_2, z_3, z_4\end{array}\right){}^T$. And if we require the wave function is normalized as
\begin{equation}\label{hole spin space}
    \langle\phi|\phi\rangle=z_1^*z_1+z_2^*z_2+z_3^*z_3+z_4^*z_4=1,
\end{equation}
then its geometry is a sphere $S^7$ in 8-d Euclid space with the symmetry group $SU(4)$. Therefore, we know that the state $|\phi\rangle$ can be mapped to a normalized five-component real vector in 5-d space $\vec{x}=(x_1,x_2,x_3,x_4,x_5)$ applying the second Hopf map $S^7\rightarrow S^4$.
Every point in $S^4$ with the components satisfying the following constraint
\begin{equation}\label{spin1}
    x_1^2+x_2^2+x_3^2+x_4^2+x_5^2=1
\end{equation}
can be the vector of hole spin direction, which has the freedom of group $SO(5)$.
 So the hole spin direction's variation can be expressed as an element in $SO(5)$ acting on the vector $\vec{x}$ just like the electron spin case. For a given $\vec{k}$, a fixed $\vec{x}$ will single out a particular direction in five-dimensional vector space, the $SO(5)$ symmetry will be broken to an $SO(4)$ symmetry. This is nothing but $SO(4)\simeq SU(2)\times SU(2)$ symmetry of the LH (light hole) and the HH (heavy hole) bands\cite{murakami20042}. As a general case, the only subspace of HH will be important thanks to the much larger heavy-hole effective mass\cite{wu2010spin}, the hole spin direction will be changed in this subspace. Since we manily focus on the conduction electron spin dynamics study in this letter, we will only discuss the electron spin in the following.
\par
\begin{figure}[h]
  \scalebox{0.4}[0.4]{\includegraphics{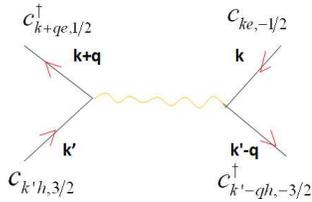}}\\
  \caption{Fenyman diagram of electron-heavy hole exchange interaction. The operators $c_{ke,-1/2}$ , $c_{k+qe,1/2}^{\dag}$ are electron with spin down annihilation  and up creation operator, respectively. Heavy hole spin up annihilation and down creation operators are $c_{k'h,3/2}$ and $c_{k'-qh,-3/2}^{\dag}$.}\label{hole}
\end{figure}
\begin{figure}[h]
  \scalebox{0.4}[0.4]{\includegraphics{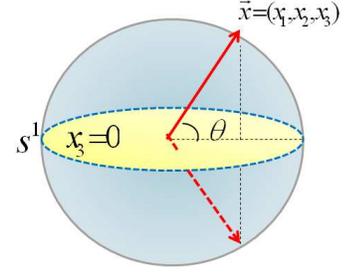}}\\
  \caption{The schematic diagram indicate the spin flip process of BAP mechanism. Through this mechanism the z component of electron spin direction is flipped. }\label{BAP}
\end{figure}

From another practical point, it will be convinient to reexpress the spin direction or denstity matrix in another representation when dealing with the many-body problems (such as electron-hole exchange terms), called second quantization representation. In fact, we can construct the one-to-one map between the annihilation and creation operators of elctron and the $i$-th componet of spin direction as follows:
\begin{equation}\label{density matrix with cc}
    x_{\vec{k},i}=c^{\dag}_{\vec{k}\sigma}(\sigma_i)_{\sigma\sigma^{'}}c_{\vec{k}\sigma^{'}},
\end{equation}
where $x_{\vec{k},i}$ is the $i$-th component of electron spin dirention at state $\vec{k}$, $c^{\dag}_{\vec{k}\sigma}, c_{\vec{k}\sigma}$ are the creation and annihilation operators at state $\vec{k}$ with spin $\sigma$, respectively. Similarly, the density matrix may be written as

 \begin{equation}\label{density cc}
    \rho=\sum_{\vec{k}}\rho_{\vec{k}}=\sum_{\vec{k}\sigma}p_{\vec{k}\sigma}c^{\dag}_{\vec{k}\sigma}c_{\vec{k}\sigma},
\end{equation}
where indices $\sigma=\uparrow, \downarrow$ in colinear space, $p_{\vec{k}\sigma}$ is the statistical weight of state $\vec{k}$ with spin $\sigma$.
Therefore, generally one can firstly analyse the physical process with anihilation/creation operators and then map them to the 3-d Eculid space using Eq.(\ref{density matrix with cc}). Applying this representation,
 the spin flip term of electron-hole exchange can be illustrated in Fig.\ref{hole}.
As can be obtained in the Fenyman diagram, that the electron spin direction will be flipped from $\uparrow$ at state $\vec{k}$ to $\downarrow$ at state $\vec{k}+\vec{q}$ or vice versa(which is not showen here). This physical picture may be described as in Fig.\ref{BAP}, after the exchange process, the vector of spin direction in the Bloch sphere will be reflected by the mirror plane $x_3=0$.
 In physics, the process may be related to spin ladder operators $\hat{S}_{\pm}=\hat{S}_x\pm i\hat{S}_y$ acting on the density matrix, but these operators are not Hermitian operators themselves, $\hat{S}_+^{\dag}=\hat{S}_-$. Thus we cannot give a Bloch sphere geometry intepretation of BAP completely here.
\subsection{various scattering in semiconductors}
\par
According to the discussions above, we know that the different scattering process  will play a significant role in electron spin dynamics. They provide additional channel for spin precession and relaxation. In this section, for completeness, we review three classic scattering that have been extensively used in transport theory in semiconductor. We write the Hamiltonian using the  second quantization representation, which will give us a clear physical process. All of the contents discussed in this section can be found in Ref.\cite{haug2008quantum,haug2004quantum}.

\begin{enumerate}
  \item  \textcolor[rgb]{0.00,0.00,0.00}{\emph{electron-phonon scattering}}
  \par A general Hamiltonian of electron-phonon interaction can be written as:
  \begin{equation}\label{electron-phonon}
    H_{ep}=\sum_{\vec{k},\vec{q}}\hbar g_{\vec{q}}c_{\vec{k}+\vec{q}}^{\dag}c_{\vec{k}}(b_{\vec{q}}+b_{-\vec{q}}^{\dag})
    \end{equation}
  where $c_{\vec{k}+\vec{q}}^{\dag}$ , $c_{\vec{k}}$ are the fermionic electron creation and annihilation operators, $b_{\vec{q}}$, $b_{\vec{q}}^{\dag}$ are the boson operators of the phonons, $g_{\vec{q}}$ is the interaction matrix element. Graphically, the electron-phonon interaction is represented by a vertex as in Fig.\ref{electron-phonon}.
\begin{figure}[h]
  \scalebox{0.5}[0.5]{\includegraphics{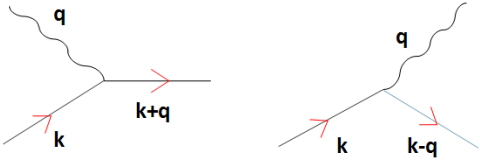}}\\
  \caption{Electron-phonon interaction. Left: A phonon (wavy line) is absorbed, while an electron (solid line) is scattered from state $\vec{k}$ to state $\vec{k}+\vec{q}$, which is represented as $ c_{\vec{k}+\vec{q}}^{\dag}c_{\vec{k}}b_{\vec{q}}$. Right: phonon emission process $ c_{\vec{k}-\vec{q}}^{\dag}c_{\vec{k}}b_{\vec{q}}$. The vertex is the interaction $g_{\vec{q}}$.}\label{electron-phonon}
\end{figure}

  \item  \textcolor[rgb]{0.00,0.00,0.00}{\emph{electron-electron scattering}}
  \par
  The electron-electron Coumlomb scattering in many-body physics can be written as:
  \begin{equation}\label{e-e}
    H_{ee}=\sum_{\vec{q},\vec{k},\vec{k'}}V_{\vec{q}}c_{\vec{k'}+\vec{q}}^{\dag}
    c_{\vec{k}-\vec{q}}^{\dag}c_{\vec{k}}c_{\vec{k'}}
  \end{equation}
  where $V_{\vec{q}}$ is the screened Coulomb potential in the random-phase approximation\cite{haug2004quantum}. The Fenyman diagram clearly describes the scattering process as in Fig.\ref{electron-electron}.
\begin{figure}[h]
  \scalebox{0.5}[0.5]{\includegraphics{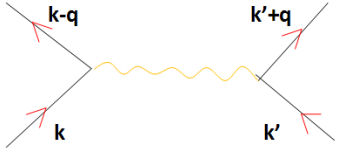}}\\
  \caption{Electron-electron interaction. Two electrons (solid line) from states $\vec{k}$ and $\vec{k'}$ through the Coumlomb scatterring (wavy line) to the final states $\vec{k}-\vec{q}$ ,$\vec{k'}+\vec{q}$, which can be described by the second quantization method such as $c_{\vec{k'}+\vec{q}}^{\dag}
    c_{\vec{k}-\vec{q}}^{\dag}c_{\vec{k}}c_{\vec{k'}}$. }\label{electron-electron}
\end{figure}

  \item \textcolor[rgb]{0.00,0.00,0.00}{\emph{electron-impurity scattering}}
  \par
  Another common scattering process in semiconductor is impurity scattering. Elastic impurity system will influence the distribution of electron in the momentum space. The interaction Hamiltonian reads
  \begin{equation}\label{e-i}
    H_{ei}=\sum_{\vec{q},\vec{k}}V_{i}(\vec{q})\rho_i(\vec{q})c_{\vec{k}+\vec{q}}^{\dag}c_{\vec{k}}
  \end{equation}
  where $V_{i}(\vec{q})$ is the electron-impurity interaction potential with the random-phase approximation, $\rho_i(\vec{q})$ is nothing but the density of the impurities in momentum space. This physical meaning is presented in Fig.\ref{e-impurity}.
  \begin{figure}[h]
  \scalebox{0.4}[0.4]{\includegraphics{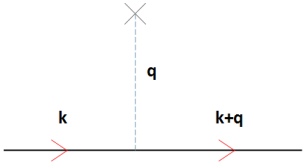}}\\
  \caption{The basic process of electron-impurity interaction. An electron (solid line) from state $\vec{k}$ is scattered to state $\vec{k}+\vec{q}$. The dotted line stands for the impurity potential. Momentum is conserved at each vertex.
     }\label{e-impurity}
\end{figure}
\end{enumerate}
\subsection{The physical interpretation of KSBE}
\par

 Based on the discussion talked above, we can generally write the dynamic equation for state of $\vec{k}$ as following\cite{jiang2009electron,wu2010spin}:
 \begin{equation}\label{general ksbe}
    \partial_t\rho_{\vec{k}}=\partial_t\rho_{\vec{k}}|_{coh}+\partial_t\rho_{\vec{k}}|_{scat}
\end{equation}
where the term $\partial_t\rho_{\vec{k}}|_{coh}$ describe the coherent precession induced by the external magnetic field or randomly distributed effective magnetic field induced by Dresshaules and Rashaba spin orbit coupling, while the scattering term $\partial_t\rho_{\vec{k}}|_{scat}$ depicts all kinds of scattering processes.
The coherence term of Eq.(\ref{general ksbe}) describes the spin dynamics during the interval of scattering.  It's during this time, the electron lives in a non interacting system. So this process can be interpreted as a Larmor procession process in a magnetic field or equivalent magnetic field  as in Fig.\ref{larmor}. The second part will mainly change the statistical weights of the corresponding states for spin-independent scattering, such as the inhomogeneous broadening mechanism. And the EY spin mechanism will be contained when in the narrow gap semiconductors, we need to insert a spin flip matrix when the scattering come up. The picture of this additional spin flip can be seen in Fig.\ref{EY}. Finally, BAP spin dynamic mechanism will need to be considered when in $p$-type semiconductors. This term will also be in the second part of Eq.(\ref{general ksbe}) due to it's a exchange interaction. The physical picture also can be understood in Fig.\ref{BAP}, where the hole spin direction will affect the electron spin direction.

\par
The coherent term in KSBE is given by
\begin{equation}
    \partial_t\rho_{\vec{k}}|_{coh}=-i[\Omega(\vec{k})\cdot\vec{\sigma}/2,\rho_{\vec{k}}]\nonumber
\end{equation}
this term is nothing but the DP mechanism like term,  the spins precess in a random magnetic field. Here, $\Omega({\vec{k}})$ may contain  Dresselhaus, Rashba or strain induced spin-orbit coupling  term\cite{wu2010spin}.
The scattering term $\partial_t\rho_{\vec{k}}|_{scat}$ contains the contribution from electron-impurity $\partial_t\rho_{\vec{k}}|_{ei}$, the electron-phonon scattering $\partial_t\rho_{\vec{k}}|_{ep}$, the electron-electron scattering $\partial_t\rho_{\vec{k}}|_{ee}$, the electron-hole Coulomb scattering $\partial_t\rho_{\vec{k}}|_{eh}$, the electron-hole exchange scattering $\partial_t\rho_{\vec{k}}|_{ex}$,
\begin{eqnarray*}
    \partial_t\rho_{\vec{k}}|_{scat}=&&\partial_t\rho_{\vec{k}}|_{ei}+\partial_t\rho_{\vec{k}}|_{ep}+
    \partial_t\rho_{\vec{k}}|_{ee}\\
    &&+\partial_t\rho_{\vec{k}}|_{eh}+\partial_t\rho_{\vec{k}}|_{ex}.\nonumber
\end{eqnarray*}
 We now give a detail description about the first term,
\begin{eqnarray*}
    \partial_t\rho_{\vec{k}}|_{ei}=&&-\pi\sum_{\vec{k'}}n_i Z_i V_{\vec{k}-\vec{k'}}^{2}\delta(\varepsilon_{\vec{k'}}-\varepsilon_{\vec{k}})(\Lambda_{\vec{k},\vec{k'}}\rho_{\vec{k'}}^{>}\Lambda_{\vec{k'},\vec{k}}\\
    &&\times \rho_{\vec{k}}^{<}-\Lambda_{\vec{k},\vec{k'}}\rho_{\vec{k'}}^{<}\Lambda_{\vec{k'},\vec{k}}\rho_{\vec{k}}^{>})+h.c.,
\end{eqnarray*}
where $n_i$ is the impurity density, $Z_i$ is the charge number of the impurity, $\varepsilon_{\vec{k'}}$ is the energy of conduction electron at state $\vec{k'}$, $\delta$ function is the energy-conserved condition for scattering processes, which is clearly indicated in the Fenyman vertex of Fig.\ref{e-impurity}, $V_{\vec{q}}$ is the screened Coulomb potential, $\Lambda_{\vec{k'},\vec{k}}$ is the spin flip matrix which describes the EY mechanism of scattering, this matrix can be removed if the EY mechanism isn't included, then the scattering can describe the inhomogeneous broadening mechanism and so on. And $\rho_{\vec{k}}^{<}=\rho_{\vec{k}}$ is the electron density matrix with state $\vec{k}$ and statistical probability $p_{\vec{k}}$, so $\rho_{\vec{k}}^{>}=1-\rho_{\vec{k}}$ can be interpreted as the statistical density matrix when the state is empty. Here, the $\delta$ function comes from the approximation of Markov approximation process, which means that the previous scattering process doesn't affect the next one.  The sum over different states in the right hand side of equation is operable only if the scattering process satisfies the conservation of momentum and energy. Therefore, one will get a differential equation group. By solving the equation group numerically, the macroscopic physical quantities, such as the electron spin density along the $\textbf{z}$-axis, can be obtained using Eq.\ref{physical quantity}.

\par
The other scattering terms like electron-electron, electron-phonon scattering can be found in Wu's related works, they possess similar structure like electron-impurity scattering above. The readers who are interested can try to understand the explicit dynamics originally from these scattering.
\section{summary}
\par
In this paper we have presented the many-body effects on spin dynamics from a Bloch sphere geometry viewpoint.  We give the explicit unitary group for usual spin dynamics mechanism based on this picture. Furthermore, the many-body effects on spin dynamics become more clearly if applying our approach. The framework outline here is relatively simple and very easy for  physicists in experiment field to have an insight to the many body effects on spin dynamics in semiconductors.
%

\end{document}